\documentclass[twocolumn,amsmath,amssymb]{revtex4}

\usepackage[T1]{fontenc}
\usepackage{color}
\usepackage[latin1]{inputenc}
\usepackage{graphicx}
\usepackage{bm}
\usepackage{epsfig}
\usepackage{rotating}

\usepackage{dcolumn}
\usepackage{bm}

\begin{document}


\title{Effect of orientational restriction on monolayers of hard ellipsoids}
       
\author{Szabolcs Varga}
\email{vargasz@almos.uni-pannon.hu}
\affiliation{Permanent address: Institute of Physics and Mechatronics, University of Pannonia, PO Box
158, Veszpr\'em, H-8201 Hungary}

\author{Yuri Mart\'{\i}nez-Rat\'on}
\email{yuri@math.uc3m.es}
\affiliation{Grupo Interdisciplinar de Sistemas Complejos (GISC), Departamento de Matem\'aticas,
Escuela Polit\'ecnica Superior,Universidad Carlos III de Madrid, Avenida de la Universidad
30, E-28911, Legan\'es, Madrid, Spain
}

\author{Enrique Velasco}
\email{enrique.velasco@uam.es}
\affiliation{Departamento de F\'{\i}sica Te\'orica de la Materia Condensada and Instituto de Ciencia de
Materiales Nicol\'as Cabrera, Universidad Aut\'onoma de Madrid, E-28049 Madrid, Spain}

\author{Gustavo Bautista-Carbajal}
\email{gustavo.bautista@uacm.edu.mx}
\affiliation{Departamento de F\'{\i}sica, Universidad Aut\'onoma Metropolitana-Iztapalapa, 09340,
M\'exico, Distrito Federal, M\'exico, and Academia de Matem\'aticas, Universidad Aut\'onoma de
la Ciudad de M\'exico, 07160, M\'exico, D. F., Mexico
}

\author{Gerardo Odriozola}
\email{godriozo@azc.uam.mx}
\affiliation{Area de F\'{\i}sica de Procesos Irreversibles, Divisi\'on de Ciencias B\'asicas e 
Ingenier\'{\i}a,
Universidad Aut\'onoma Metropolitana-Azcapotzalco, Av. San Pablo 180, 02200 M\'exico, D.F.,
Mexico.
}
\date{\today}

\begin{abstract}
The effect of out-of-plane orientational freedom on the orientational ordering
properties of a monolayer of hard ellipsoids is studied using Parsons-Lee scaling approach
and replica exchange Monte Carlo computer simulation. Prolate and oblate ellipsoids exhibit
very different ordering properties, namely, the axes of revolution of prolate particles tend to
lean out, while those of oblate ones prefer to lean into the confining plane. The driving
mechanism of this is that the particles try to maximize the available free area on the confining
surface, which can be achieved by minimizing the cross section areas of the particles with the
plane. In the lack out-of-plane orientational freedom the monolayer of prolate particles is
identical to a two-dimensional hard ellipse system, which undergoes an isotropic-nematic
ordering transition with increasing density. With gradually switching on the out-of-plane
orientational freedom the prolate particles lean out from the confining plane and a
destabilisation of the in-plane isotropic-nematic phase transition is observed. The system of
oblate particles behaves oppositely to that of prolates. It corresponds to a two-dimensional
system of hard disks in the lack of out-of-plane freedom, while it behaves similar to that of
hard ellipses in the freely rotating case.

\end{abstract}


\maketitle

\section{Introduction}

Phase behavior of non-spherical hard bodies with their centers of mass confined in
planar geometry is receiving considerable attention due to the recent development of the
preparation of colloidal particles with various shapes and new experimental techniques.
Nowadays it is possible to prepare colloids with several geometrical shapes such as cubes,
polyhedrons, octopods, ellipsoids and helices [1-6]. The anisotropic colloids can be confined
at the interfaces [7-11], between two parallel solid walls [12,13], at the bottom of the sample
holder [14], at a substrate surface [15] and into a lamellar matrix of surfactants [16,17]. The
confinement can be so strong that even colloidal monolayers can be realized experimentally.
Ordering properties of two-dimensional and quasi two-dimensional (q2D) non-
spherical colloids has been the subject of several experimental and theoretical studies [17-36].
The reason for this is that the nature of two-dimensional (2D) nematic ordering is quite
different from the three-dimensional (3D) one. It shows only quasi-long-range orientational
order with algebraically decaying orientational correlations and the ordering transitions
between isotropic and nematic phases are first order or continuous through a Kosterlitz-Thouless
disclination unbinding type mechanism [18, 37].
Strictly 2D colloidal systems cannot be examined experimentally, because the out-of-
plane orientational and positional freedoms are always present to some extent. Therefore it is
desirable to extend the theoretical studies in such directions, where the extra orientational and
positional freedoms are present. Along this line, the ordering properties of microtubules
confined in a thin slit have been modeled as hard spherocylinders placed between two planar
walls in [28]. In agreement with the experiment it has been found that the isotropic-nematic
transition density increases with the wall separation [28].
Ellipsoidal shaped colloidal particles are gaining widespread applications due to
development of the stretching techniques in preparation of monodisperse prolate and oblate
ellipsoids from spherical latex particles [4,38]. They can be also confined into planar
geometry and create a monolayer to study ordering and glassy behavior of the q2D ellipsoid
systems [12,13].

In this paper we examine the orientational ordering properties of q2D hard ellipsoid
systems, where particles are allowed to rotate out the confinement plane to some extent, while
it is assumed in first approximation that the centers of the particles are always in the same
plane. Switching on the out-of-plane orientational freedom by gradual increase (decrease) of
the limiting polar angle ($\theta_c$) for oblate (prolate) shapes, it is possible to make a link between
strictly 2D hard ellipse (hard disk) systems for prolate (oblate) shapes and q2D freely rotating
prolate (oblate) ellipsoids systems. We show that the additional out-of-plane orientational
freedom changes substantially the orientational ordering and the transition properties of both
oblate and prolate shaped ellipsoids. To maximize the available free area on the confining
surface, the axis of revolution of the prolate particle leans out from the plane, while the oblate
particle leans into the plane. As a consequence, the freely rotating prolate ellipsoids resemble
hard disks at high densities, while those of oblate ellipsoids behave similarly to hard ellipses.
The isotropic-nematic phase transition of hard ellipses corresponds to a planar nematic-biaxial
nematic phase transition of the oblate ellipsoids. Here we note that our present work can be
considered as an extension of our previous studies done for monolayers of uniaxial and
biaxial hard particles using restricted-orientation approximation [39,40].

The paper is organized as follows. The molecular model and the details of the
confinement are presented in Sec. II. Sec. III is devoted to the Parsons-Lee theory of q2D hard
ellipsoid fluids, where we show how to determine the equilibrium free energy, surface
coverage, and order parameters of the uniaxial and biaxial nematic phases. Technical details
of the replica exchange Monte Carlo simulation method are given in Sec. IV. The order
parameters, the surface coverage, the equation of state and the phase diagram of the system
hard ellipsoids are presented in Sec. V. Finally, some conclusions are drawn in Sec. VI.

\section{Molecular model}

\begin{figure}
\begin{center}
\includegraphics[width=0.9\linewidth,angle=0]{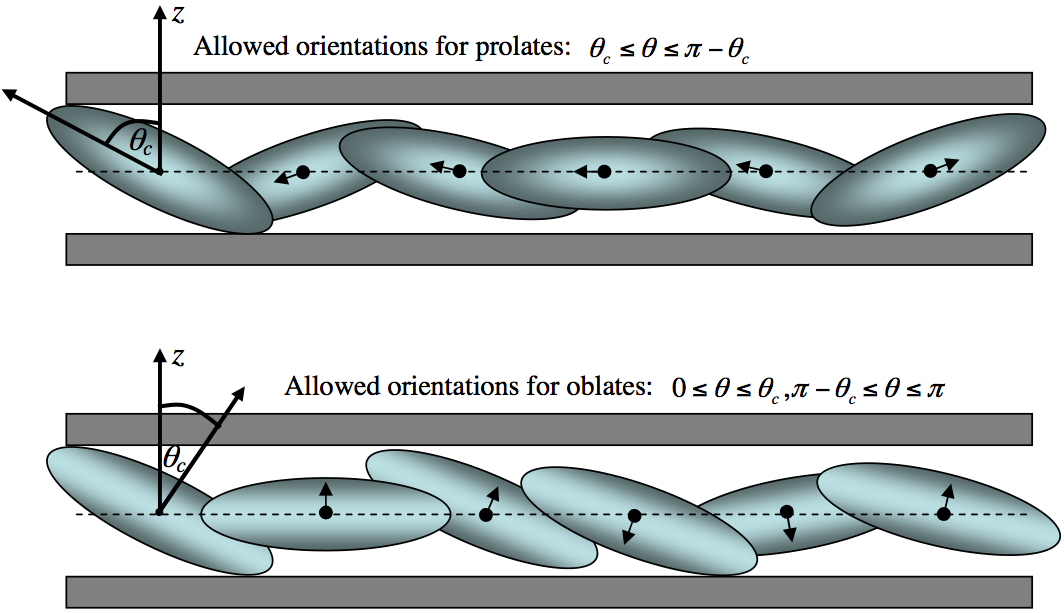}
\caption{
Schematic representation of the system of confined hard ellipsoids: prolate
ellipsoids (upper panel), oblate ellipsoids (lower panel). The particles are confined both
positionally and orientationally. The center of mass of the ellipsoid is always in the XY plane,
which is denoted by the dashed line, while the ellipsoid can rotate freely in azimuthal angle
($\phi$) and is restricted by the polar angle ($\theta_c$). The arrows indicate the direction of the main symmetry axis of the ellipsoids.
}
\label{fig1}
\end{center}
\end{figure}

In this work we examine the q2D system of hard ellipsoids, where the particles are
allowed to rotate freely in a restricted region of the solid angle, while the centers of particles
can move only on a two-dimensional XY plane. The shape of the particles can be both prolate
and oblate, i.e. the aspect ratio or shape anisotropy ( $\kappa=\sigma_{\parallel}/\sigma_{\perp}$, 
where $\sigma_{\parallel}$ and $\sigma_{\perp}$  are the
lengths along and perpendicular to the axis of revolution of the ellipsoid, respectively) can be
either larger than one ($\kappa>1$ , prolate shape) or between zero and one ($0<\kappa<1$, oblate
shape). The orientational restriction takes place only in the polar angle ($\theta$), which is measured
from the axis of revolution of each particle to the normal of the confining XY plane. While
the particles can rotate freely in the azimuthal angle ($0<\phi<2\pi$), the allowed range of the
polar angle is different for prolate and oblate ellipsoids, namely, the range of the polar angle
is given by $\theta_c<\theta<\pi-\theta_c$ for prolate shapes, while $0<\theta<\theta_c$ and 
$\pi-\theta_c<\theta<\pi$ intervals
are allowed for oblate ones (see Fig. \ref{fig1}). We can tune $\theta_c$ between 0 and $\pi/2$. In the case of
prolate ellipsoids ($\kappa>1$) $\theta_c=0$ corresponds to the confined system of freely rotating
particles, while $\theta_c=\pi/2$ gives the two-dimensional system of hard ellipses. The situation is
different for oblate ellipsoids ($0<\kappa<1$), because the hard disk limit is given by $\theta_c=0$ , while
the freely rotating system of confined oblate ellipsoids corresponds to $\theta_c=\pi/2$. The
interaction between ellipsoids is purely hard, i.e. the pair potential between two particles is
given by
\begin{eqnarray}
u\left(r_{12},\boldsymbol{\omega}_{12},\boldsymbol{\omega}_1,\boldsymbol{\omega}_2\right)=
\left\{
\begin{matrix}
\infty, & \text{for} \ r_{12}\leq \sigma(\boldsymbol{\omega}_{12},\boldsymbol{\omega}_1,\boldsymbol{\omega}_2),\\
0, & \text{otherwise},
\end{matrix}
\right.
\end{eqnarray}
where $r_{12}$ is the center-to-center distance, $\boldsymbol{\omega}_{12}=
(\cos\phi_{12},\sin\phi_{12},0)$ is the unit vector connecting
the centers of the two ellipsoids, $\boldsymbol{\omega}_i=
(\sin\theta_i\cos\phi_i,\sin\theta_i\sin\phi_i,\cos\theta_i)$ is the orientational unit
vector of particle $i$ ($i=1,2$) and $\sigma(\cdots)$ is the distance of closest approach. 
Note that the third component of $\boldsymbol{\omega}_{12}$ is always zero, which ensures 
that the centers of the particles are always in
the XY plane. The distance of closest approach between two ellipsoids [$\sigma(\cdots)$] is 
approximated in the manner proposed in our previous works [41,42].

\section{Parsons-Lee theory}
To describe theoretically the orientational ordering properties of the monolayer of hard
ellipsoids we derive our working equations from the well-known Parsons-Lee (PL) theory of
hard bodies [43,44], which proved very successful in determination of the equation of state
and the transition properties of IN phase coexistence of non-spherical hard body fluids both in
two [23,24] and three dimensions [45-47]. Here we present only the important ingredients of
the theory, specific equations for the ellipsoid monolayer and some technical details.
It is customary to deal with the free energy of the system, which is the sum of ideal
and excess free energy terms, i.e. $F=F_{\rm id}+F_{\rm ex}$ . The ideal term can be determined 
exactly from
\begin{eqnarray}
\frac{\beta F_{\rm id}}{N}=\log \rho -1+\int d\omega f(\omega) 
\log f(\omega),
\end{eqnarray}
where $\beta=1/k_{\rm B} T$ is the inverse temperature, $N$ is the number of particles, 
$\rho=N/A$ , $A$ is the area of the plane, $\omega=(\phi,\theta)$ is the collection azimuthal 
and polar angles, $d\omega=d\phi d\theta \sin\theta$ and
$f(\omega)$ is the normalized orientational distribution function
($\int d\omega f(\omega)=1$). The ranges of the azimuthal and polar angles in the integrations 
have been already given in Sec. II. The excess
free energy contribution can be obtained approximately with the mapping procedure from the
actual system to a reference one, where the second virial coefficient and the excess free
energy can be obtained with good accuracy. In our case we choose the system of 2D hard
disks as a reference system, because our confined ellipsoid system is q2D. The second virial
coefficient of the hard disks is given by $B_2^{\rm HD}=D^2\pi/2$, where $D$ is the diameter 
of the hard disk. A luckily simple and accurate expression can be derived for the excess free 
energy of hard disks, too, using the scaled particle theory [48]
\begin{eqnarray}
\frac{\beta F_{\rm ex}^{\rm HD}}{N}=-\ln\left(1-\eta_{\rm HD}\right)+
\frac{\eta_{\rm HD}}{1-\eta_{\rm HD}},
\end{eqnarray}
where $\eta_{\rm HD}=\rho a_{\rm HD}$ is the packing fraction of the hard disk and 
$a_{\rm HD}=B_2^{\rm HD}/2$ is the area of
the hard disk. To perform the mapping procedure from the hard ellipsoids into hard disks we
introduce the surface coverage (or packing fraction) of the hard ellipsoids, because the
intersection of the ellipsoid with the XY plane is an ellipse with characteristic lengths
depending on the orientation of the ellipsoid. These characteristic lengths of the ellipse can be
calculated easily using simple geometry. One can derive that $\sigma_{\perp,e}=\sigma_{\perp}$ and
$\sigma_{\parallel,e}=\sigma_{\parallel}\sigma_{\perp}/\sqrt{\sigma_{\perp}^2\sin^2\theta
+\sigma_{\parallel}^2\cos^2\theta}$,
where $\theta$ is the polar angle of the ellipsoid. This means
that the monolayer of the hard ellipsoids can be visualized as a multicomponent mixture of
hard ellipses on the XY plane. Using the dimensions of the hard ellipse, the intersected area of
the hard ellipsoid with the XY plane can be obtained from
\begin{eqnarray}
a(\theta)=\frac{\pi\sigma_{\perp}^2\sigma_{\parallel}}{4\sqrt{\sigma_{\perp}^2\sin^2\theta+
\sigma_{\parallel}^2\cos^2\theta}}.
\end{eqnarray}
This polar angle dependent area together with the orientational distribution function allows us
to determine the surface coverage of the plane by ellipses, which is given by 
$\eta=\rho \langle a\rangle$,
where the average area of the intersected ellipse can be determined from 
$\langle a\rangle=\int d\omega a(\theta)f(\omega)$.
Note that the surface coverage gives back the packing fraction of hard disks in the hard-sphere
limit, because $\sigma_{\parallel}=\sigma_{\perp}=D$ and the distribution function is a constant. 
In PL approach the
area of hard disks and the average area of hard ellipses are assumed to be the same
($a_{\rm HD}=\langle a\rangle$), which implies also that $\eta_{\rm HD}=\eta$ due to one-to-one 
correspondence between the
two systems. Based on the concept of Parsons and Lee we can now calculate the excess free
energy of the monolayer of hard ellipsoids using the thermodynamic properties of hard disks
as follows
\begin{eqnarray}
\frac{\beta F_{\rm ex}}{N}\approx \frac{\beta F_{\rm ex}}{N}\frac{B_2^{\rm HE}}{B_2^{\rm HD}},
\end{eqnarray}
where the second virial coefficient of hard ellipsoids restricted to XY plane is given by
\begin{eqnarray}
B_2^{\rm HE}=\frac{1}{2}\int d\omega_1 f(\omega_1)\int d\omega_2 f(\omega_2) 
A_{\rm excl}(\omega_1,\omega_2).
\end{eqnarray}
This equation contains the excluded area between two ellipsoids, which can be obtained from
the distance of closest approach ($\sigma$) as follows
\begin{eqnarray}
A_{\rm excl}(\omega_1,\omega_2)=\frac{1}{2}\int_0^{2\pi}d\phi_{12}\sigma^2(
\boldsymbol{\omega}_1,\boldsymbol{\omega}_2,\boldsymbol{\omega}_{12}).
\end{eqnarray}
Using Eqs. (3)-(6) and $\eta_{\rm HD}=\eta$ condition the excess free energy of the system becomes
\begin{eqnarray}
&&\frac{\beta F_{\rm ex}}{N}=\left[-\ln(1-\eta)+\frac{\eta}{1-\eta}\right]\nonumber\\ 
&&\times\frac{\int d\omega_1f(\omega_1)\int d\omega_2f(\omega_2)A_{\rm excl}(\omega_1,\omega_2)}
{4\int d\omega a(\theta)f(\omega)}.
\end{eqnarray}
The sum of Eqs. (2) and (8) constitutes our density functional equation to determine the
equilibrium orientational distribution function [$f(\omega)$] and the total free energy density at
a given density ($\rho=N/A$). The minimization of the free energy functional with respect
to $f(\omega)$ gives us the Euler-Lagrange equation for the equilibrium $f(\omega)$. Note that the
minimization must be carried out by maintaining the normalization condition
($\int d\omega f(\omega)=1$) . Although the minimization procedure is very simple and straightforward,
we do not present the equation of $f(\omega)$ , since the resulting equation is too long. Once the
equilibrium $f(\omega)$ is determined with a standard iterative method, the free energy density
can be obtained by the substitution of the resulting $f(\omega)$ into the free energy functional
(sum of Eq. (2) and (8)).
We determine the pressure from the free energy using
$P=\rho^2\partial(F/N)/\partial\rho$ and examine the orientational ordering properties of 
the monolayer by
the standard uniaxial and biaxial order parameters. These are defined as
\begin{eqnarray}
S=\langle P_2\rangle=\int d\omega f(\omega)P_2(\theta),
\end{eqnarray}
and
\begin{eqnarray}
\Delta=\int d\omega f(\omega)\sin^2\theta \cos(2\phi),
\end{eqnarray}
where $P_2(x)=3x^2/2-1/2$ is the second order Legendre polynomial. The uniaxial order
parameter ($S$) is positive for out-of-plane ordering ($0<S<1$), while it is negative for in-
plane ordering ($-1/2<S<0$). The biaxial order parameter ($\Delta$) can be very useful in finding
in-plane orientational ordering transitions, because $\Delta$ is zero for in-plane complete
disorder, while it is nonzero for in-plane order. $S$ is practically a three-dimensional order
parameter of bulk ellipsoids, while $\Delta$ is the corresponding two-dimensional one of bulk
ellipses when $\theta_c=\pi/2$ . Since our system is q2D, we need both of them. To see the
effect of out-of-plane orientational freedom and to which extent the particles are tilted off
from the confining plane, we determine the average aspect ratio of the effective ellipse
system, which is defined as $\kappa_{\rm eff}=\langle\sigma_{\parallel,e}/\sigma_{\perp,e}\rangle$.
In Sec. V we plot the density and the pressure in dimensionless units, which are defined as 
$\rho^*=N\sigma_{\perp}^2/A$ and $P^*=\beta P\sigma_{\perp}^2$.

\section{Replica Exchange Monte Carlo simulation}

As for the 2D hard ellipse study [35], we are implementing the replica exchange
Monte Carlo technique [49-51]. This is done to avoid, as far as possible, the inherent
hysteresis associated to transitions [52]. The method is based on the definition of an extended
ensemble with partition function $Q_{\rm ext}=\prod_{i=1}^{n_r} Q_i$, where $Q_i$ is the partition 
function of ensemble $i$. $n_r$ ensembles are considered, and $n_r$ replicas are employed to sample the extended
ensemble, each one at each ensemble. Defining $Q_{\rm ext}$ allows introducing swap trial moves
between any two replicas, whenever the detailed balance condition is satisfied. In our case it
is convenient to expand isobaric-isothermal ensembles in pressure [53]. This is so since we
are studying hard particles. Hence, the partition function of the extended ensemble reads
[53,54]
\begin{eqnarray}
Q_{\rm ext}=\prod_{i=1}^{n_r} Q_{N T P_i},
\end{eqnarray}
where $Q_{NTP_i}$ is the partition function of the isobaric-isothermal ensemble of the system at
pressure $P_i$ and temperature $T$. $N$ particles are taken into account at each ensemble. A
standard implementation is used to sample the NTP i ensembles. This implies independent trial
2D displacements, 3D rotations of single ellipsoids, and area changes of the simulation cell. In
case of having confining planes, 3D rotations of single ellipsoids are constrained by their
presence. We are also accounting for non-orthogonal parallelogram cells and so, additional
trial changes of the angles and relative length sides of the cell lattice vectors are included. The
following acceptance rule is set [53]
\begin{eqnarray}
P_{\rm rm,acc}=\text{min}\left\{1,\exp\left[\beta\left(P_i-P_j\right)\left(A_i-A_j\right)\right]
\right\},
\end{eqnarray}
where $A_i-A_j$ is the area difference between replicas $i$ and $j$ . Adjacent pressures must be
close to provide swap acceptance rates over 0.1. Simulations are started from a packed
triangular arrangement of spheres which are elongated in the direction normal to the plane by
a factor $\kappa$ . In case of oblates, a stretching factor $\kappa$ is also applied in a certain 
in-plane direction. Conversely to the stretching of spheres in a 3D cell, this procedure leads to the
largest packed arrangement of spheroids in a plane [55]. A stationary state is reached faster by 
decompressing packed cells than by compressing lose random configurations [52]. We
perform the necessary trial moves to observe a stationary state. At this stage we adjust
maximum displacements to produce acceptance rates close to 0.3. We also relocate all
pressures, initially set by following a geometric progression with the replica index, to obtain
similar swap acceptance rates for all pairs of adjacent ensembles. Next, we perform $4\times 10^{12}$
sampling trials for fixed maximum particle displacements, maximum rotational
displacements, maximum area changes, and maximum changes of the lattice vectors. Verlet
neighbor lists [56] are used to improve performance. We set $N\sim 400$ ellipsoids and $n_r$ as a
function of the pressure range to be covered. $N\sim 400$ is sufficiently large in view of Xu et. al.
analysis of system size effects [33]. More details on the employed methods are given in
previous works [52].

\section{Results and discussion}

\begin{figure}
\begin{center}
\includegraphics[width=0.8\linewidth,angle=0]{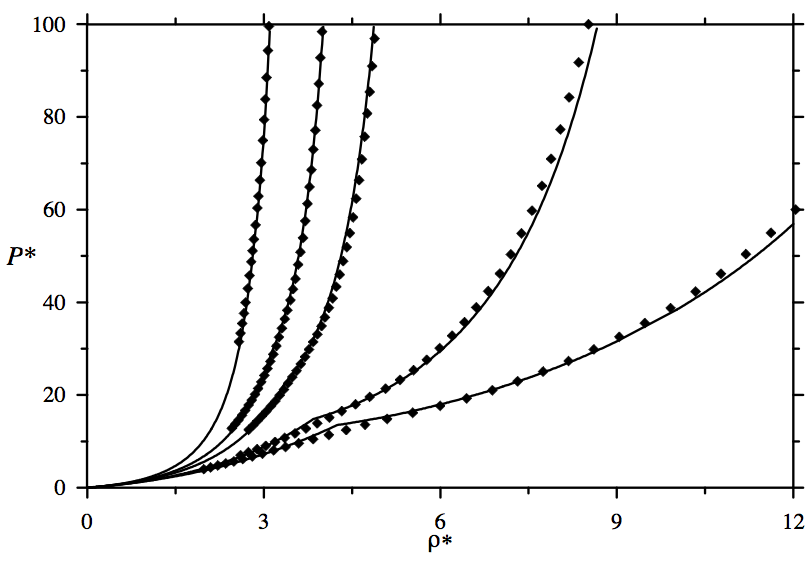}
\includegraphics[width=0.8\linewidth,angle=0]{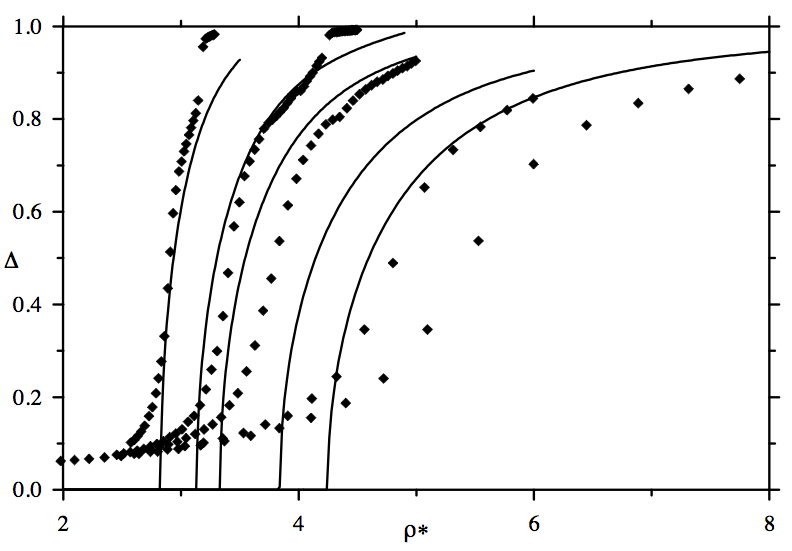}
\caption{
Orientational ordering of the system of hard ellipses: the equation of state (upper
panel) and the two-dimensional order parameter vs. density (lower panel). The curves
correspond to $\kappa^{-1}=3$, 4, 5, 10 and 20 from left to right. The curves are the results 
of PL theory,
while the diamond symbols correspond to MC simulation results. The hard ellipse system is
obtained from oblate ellipsoids by tilting their orientations into the XY plane.
}
\label{fig2}
\end{center}
\end{figure}

In this section we present our theoretical and simulation results for hard ellipses, freely
rotating and orientationally restricted monolayers of hard ellipsoids. The equation of state, the
surface coverage, the in-plane and out-of-plane order parameters and the isotropic-nematic
(IN) transition densities are determined for various values of shape anisotropy. We start with
the system of hard ellipses, which is obtained by setting the polar angles of all ellipsoidal
particles ($\theta$) to be $\pi/2$. This condition is accompanied by $f(\omega)=f(\phi)$ and that 
the uniaxial order parameter ($S$) is always -1/2 and the biaxial order parameter can be obtained from
$\Delta=\int_0^{2\pi}d\phi f(\phi)\cos(2\phi)$. Note that $\Delta$ now serves as a 2D orientational 
order parameter. The
theoretical calculations and simulations can also be carried out with both oblate-shaped
($0<\kappa<1$) and prolate-shaped hard ellipsoids ($\kappa>1$), where $\theta=\pi/2$ for all particles. Here we
perform the hard ellipse study using oblate-shaped ellipsoids. Fig. \ref{fig2} shows the pressure and
the 2D order parameter ($\Delta$) as a function of density for $\kappa^{-1}=3$, 4, 5, 10 and 20. 
One can see
that more particles on the surface (higher densities) are required to maintain the same value of
the pressure with increasing shape anisotropy. This is due to the fact that the XY plane
becomes more spacious with increasing $\kappa^{-1}$ , i.e. $\sigma_{\parallel}$ must decrease 
at fixed $\sigma_{\perp}$ . 

\begin{figure}
\begin{center}
\includegraphics[width=0.9\linewidth,angle=0]{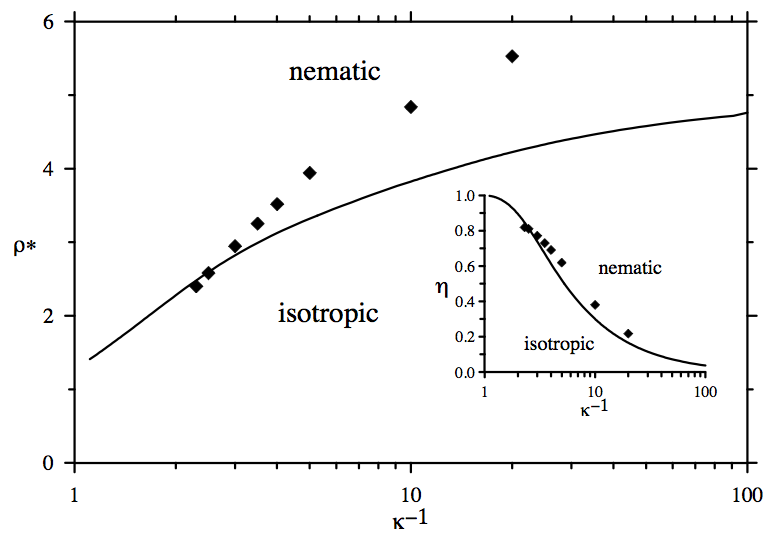}
\caption{
Shape dependence of the isotropic-nematic (IN) transition of hard ellipses: reduced
density vs. aspect ratio. The curves are the results of PL theory, while the diamond symbols
correspond to MC simulation results. The inset shows the packing fraction of the IN transition
as a function of aspect ratio, where $\eta=\rho\pi\sigma_{\parallel}\sigma_{\perp}/4$.
}
\label{fig3}
\end{center}
\end{figure}

It can be also
seen in Fig. \ref{fig2} that the simulation data are well reproduced by the approximate PL theory
except the values of the order parameters of the N phases corresponding to extremely
anisotropic particles ($\kappa^{-1}=10$ and 20). These data well agree with those given in [33]. The
order parameter curves reveal for the occurrence of isotropic-nematic phase transition, because
the phase is isotropic at low densities ($\Delta=0$), while it is nematic at high ones ($\Delta>0$). 
The phase
transition is second order in the theory, while it is higher order (Kosterlitz-Thouless type
continuous transition) in the simulation. Both the theory and the simulation show the same
tendencies for IN transition densities and packing fractions (see Fig. \ref{fig3}). Making the 
ellipsoids
more anisotropic one can see that the IN transition density increases, while the IN packing
fraction decreases. This apparent contradiction is due to the fact that the increasing shape
anisotropy makes the system more spacious ($\sigma_{\perp}$ is constant, while 
$\sigma_{\parallel}$ is decreasing), which
requires more particles for the initiation of the phase transition, while the XY plane can be less
occupied at the same time. It can be also seen that the theory underestimates IN transition
densities and packing fractions in higher extent with increasing shape anisotropy, which is
due to the fact that the contribution of higher order virial coefficients are not negligible with
increasing $\kappa^{-1}$.

\begin{figure}
\begin{center}
\includegraphics[width=0.8\linewidth,angle=0]{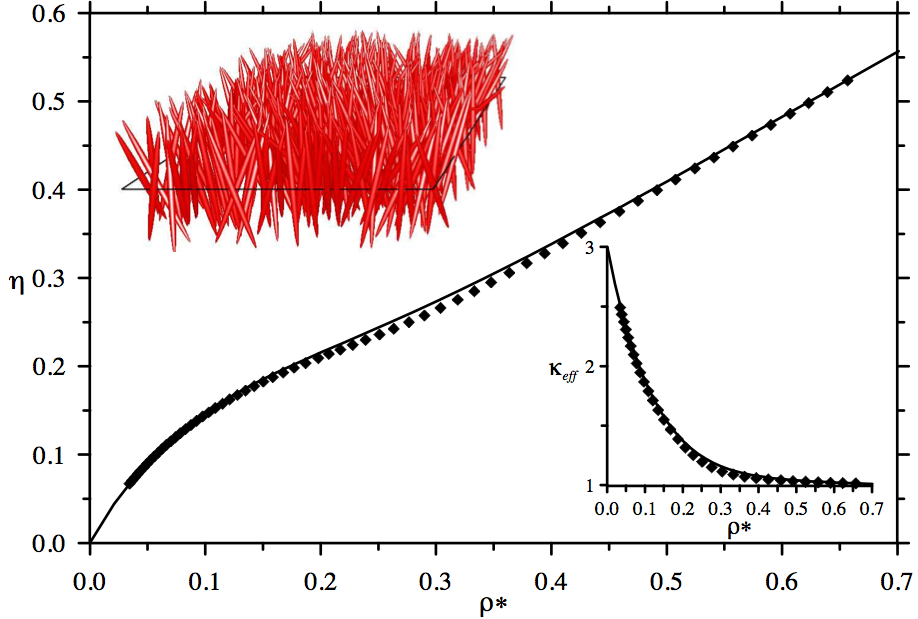}
\includegraphics[width=0.8\linewidth,angle=0]{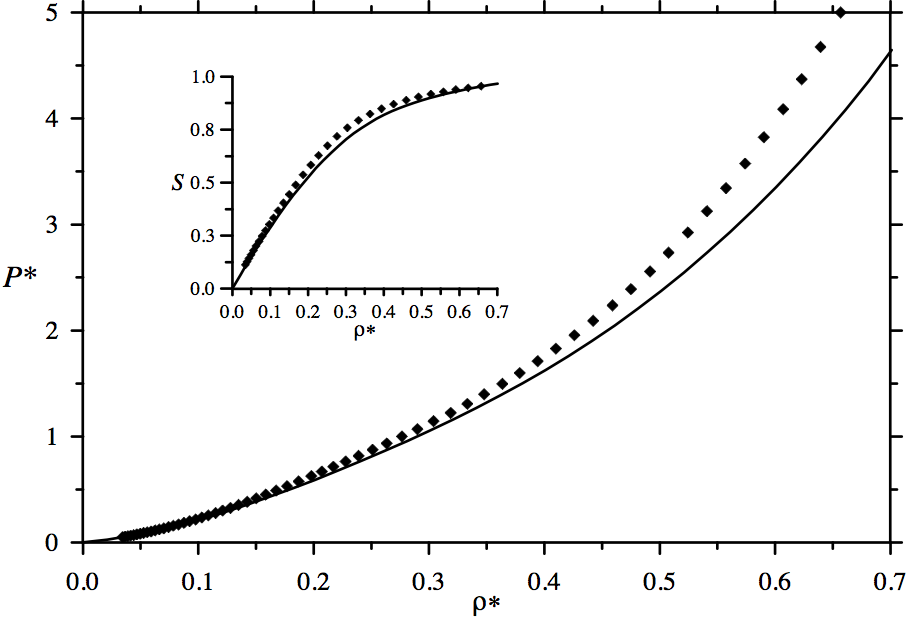}
\caption{
Monolayer of freely rotating hard prolate ellipsoids: surface coverage (packing
fraction) vs. reduced density (upper panel) and the equation of state (lower panel) at $\kappa=10$.
The effective aspect ratio of the corresponding hard ellipse system
($\kappa=\langle\sigma_{\parallel,e}/\sigma_{\perp,e}\rangle$) is shown in the inset of the 
upper panel, while
the out-of-plane order parameter ($S$) is presented in the inset of the lower panel. The curves
are the results of PL theory, while the diamond symbols correspond to MC simulation results.
The snapshot is the result of replica exchange Monte Carlo simulation method.
}
\label{fig4}
\end{center}
\end{figure}

\begin{figure}
\begin{center}
\includegraphics[width=0.8\linewidth,angle=0]{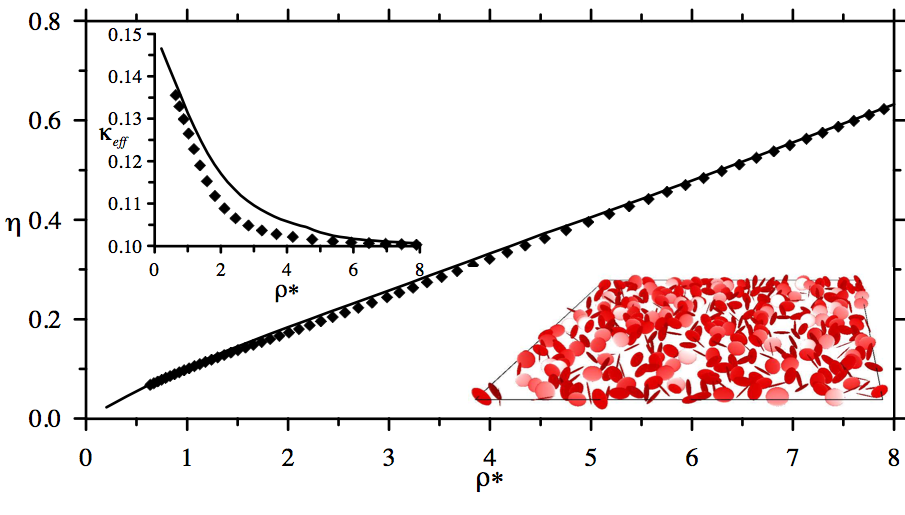}
\includegraphics[width=0.8\linewidth,angle=0]{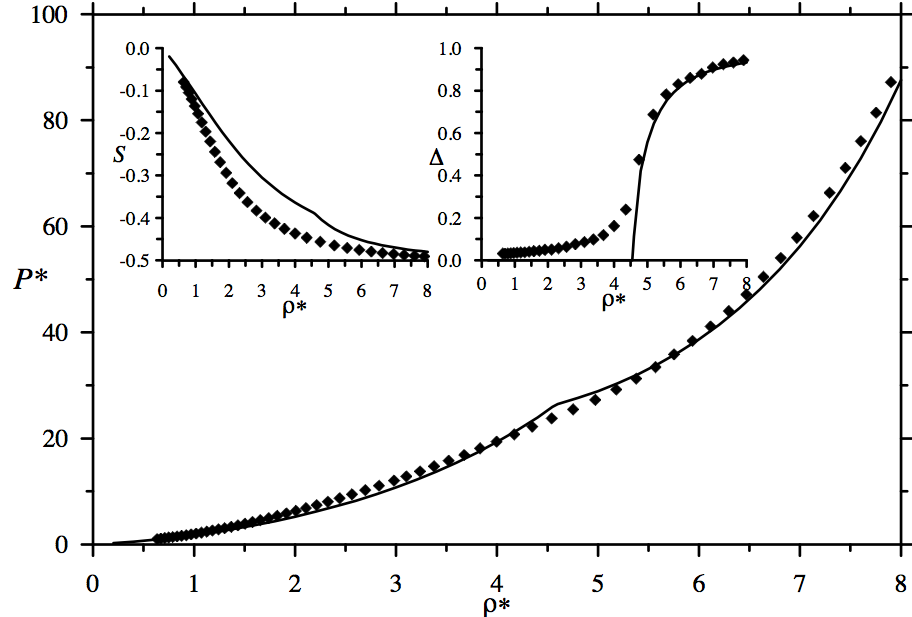}
\caption{
Monolayer of freely rotating hard oblate ellipsoid monolayer: surface coverage
(packing fraction) vs. reduced density (upper panel) and the equation of state (lower panel) at
$\kappa=1/10$. The effective aspect ratio of the corresponding hard ellipse is shown in the inset of
the upper panel, while the out-of-plane ($S$) and the in-plane ($\Delta$) order parameters are presented
in the inset of the lower panel. The curves are the results of PL theory, while the diamond
symbols correspond to MC simulation results. The snapshot is the result of replica exchange
Monte Carlo simulation method.
}
\label{fig5}
\end{center}
\end{figure}

The monolayer of freely rotating hard ellipsoids can be achieved by using $\theta_c=\pi/2$
and $\theta_c=0$ limiting angles for oblate and prolate shapes, respectively. The resulting surface
coverage, effective aspect ratio, pressure and order parameters are shown for prolate and
oblate shaped ellipsoids in Figs. \ref{fig4} and \ref{fig5}, respectively. In these figures 
the effective aspect ratio of the corresponding hard ellipse system is given by
$\kappa_{\rm eff}=\langle \sigma_{\parallel,e}/\sigma_{\perp,e}\rangle
=\langle \kappa/\sqrt{\sin^2\theta+\kappa^2\cos^2\theta}\rangle$,
which is the average aspect ratio of the ellipses
obtained by intersecting the ellipsoids by the XY plane. In turn, the surface coverage can be
obtained with the help of $\kappa_{\rm eff}$  as follows: $\eta=\rho^*\pi \kappa_{\rm eff}/4$ . 
Starting with prolates ellipsoids,
one can see that the particles do not form an isotropic phase, but they align along the normal
of the XY plane (see the simulation snapshot of Fig. \ref{fig4}) even at very low densities. 
In fact, an
isotropic phase would appear only at the limit of infinite dilution. The density dependence of
the surface coverage and the effective aspect ratio shows that the particles have less and less
chance to lie into the XY surface with increasing density because of packing effects. The
intersection of the ellipsoid with the XY surface is practically a hard disk for $\rho^*>0.6$, where
the effective ellipse aspect ratio is almost one. This makes the monolayer of hard ellipsoids
very similar to the system of hard disks at high densities despite the presence of orientational
fluctuations. The in-plane order parameter (Eq. 10) is always zero, while the out-of-plane
order parameter (Eq. 9) is positive. This shows that the out-of-plane orientational ordering is
uniaxial with nematic director parallel to the layer normal. It can be also seen that $S$ is very
close to its maximum value ($S_{\rm max}=1$) at $\rho^*=0.7$, which corresponds to the case where all
prolate ellipsoids are parallel and align along the layer normal. The reason why prolate
ellipsoids prefer the out-of-plane ordering is that they can maximize the free area available on
the surface with the penalty of orientational entropy loss. Moreover the close packing
structure of prolate ellipsoids is identical to that of 2D hard disks. One can also see that the
theory reproduces quite well the simulation data for all properties except the equation of state
at high densities. Regarding the possible freezing transition at high densities, it has not been
examined here, because it is not the scope of our present study.

The principle of minimizing the intersected area with the XY plane applies also for
oblate ellipsoids (see Fig. \ref{fig5}). The minimal intersected area, which is actually an 
ellipse with
$\sigma_{\parallel,e}=\sigma_{\parallel}$ and $\sigma_{\perp,e}=\sigma_{\perp}$ dimensions, 
can be achieved with ordering into the XY plane, while
the out-plane ordering result in higher intersected area, because 
$\sigma_{\parallel}<\sigma_{\perp}$ for oblate ellipsoids.
The resulting in-plane ordering can be seen in the simulation snapshot and from the high
density behavior of the effective aspect ratio ($\kappa_{\rm eff}\to\kappa=\sigma_{\parallel}/
\sigma_{\perp}$). The out of plane order
parameter goes to -1/2, which corresponds to complete in-plane ordering. This in-plane order
is isotropic at low densities ($\Delta=0$), while it is nematic at high densities ($\Delta>0$). 
This means
that the system undergoes a phase transition from a planar uniaxial nematic order
($S<0$, $\Delta=0$) to a biaxial nematic one ($S<0$, $\Delta\neq 0$), which can be considered as a 
2D IN phase transition. The order of the uniaxial nematic-biaxial nematic (N-BN) phase transition is
proved to be second order in the approximate PL theory, while the transition is higher order
continuous in the simulation. This transition is the result of the competition between in-plane
orientational entropy (favoring disorder) and the in-plane packing entropy (favoring order).
The high density structure of the hard ellipsoid monolayer resembles the nematic phase of the
2D hard ellipses, while at low densities the structure more or less corresponds to an isotropic
phase of a polydisperse mixture of hard ellipses. One can also see that the agreement between
the theory and the simulation is quite good for oblate ellipsoids, especially with respect to the
equation of state. An interesting feature is that the theory overestimates the out-of-plane
ordering; this could stem from the effect of higher virial coefficients, which are not included
in the theory.

\begin{figure}
\begin{center}
\includegraphics[width=0.9\linewidth,angle=0]{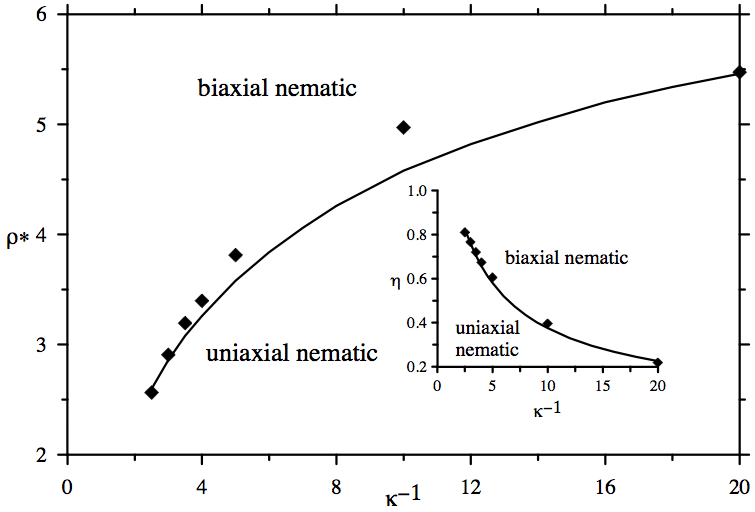}
\caption{
Shape dependence of the uniaxial nematic-biaxial nematic (N-BN) transition of
freely rotating hard oblate monolayer: reduced density vs. aspect ratio. The curves are the
results of PL theory, while the diamond symbols correspond to MC simulation results. The
inset shows the packing fraction of the N-BN transition as a function of aspect ratio.
}
\label{fig6}
\end{center}
\end{figure}

\begin{figure}
\begin{center}
\includegraphics[width=0.9\linewidth,angle=0]{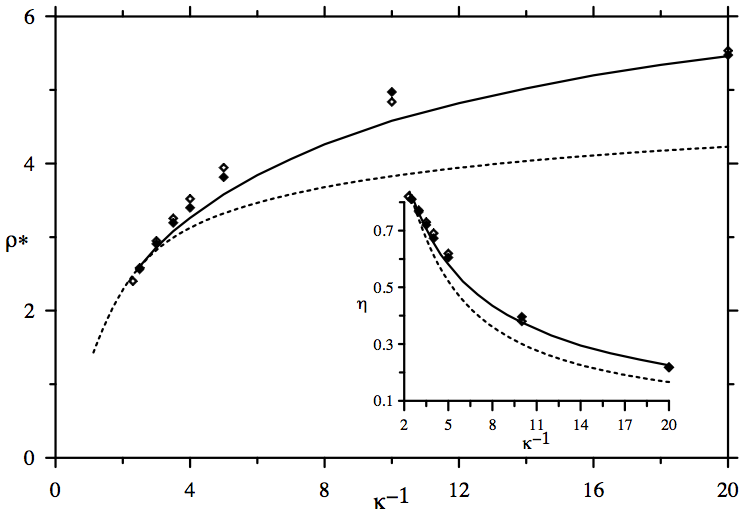}
\caption{
Comparison of the transition densities of freely rotating hard ellipses and confined
hard oblates in the reduced density-aspect ratio plane. The continuous (ellipsoids) and dashed
(ellipses) curves are the results of PL theory, while the open and filled diamond symbols
correspond to MC simulation results for hard ellipses and hard ellipsoids, respectively. The
inset shows the packing fractions of IN and N-BN transitions as a function of the aspect ratio.
}
\label{fig7}
\end{center}
\end{figure}

Fig. \ref{fig6} shows the in-plane isotropic-nematic (or N-BN) transition densities 
as a function of aspect ratio resulting from the PL theory and simulation. On the other hand, the
transition density increases with increasing shape anisotropy, while the packing fraction
decreases. This is due to the fact that we go to the "volumeless" hard needle limit with
decreasing $\kappa$ , where the transition density saturates at finite value, while the packing 
fraction vanishes (note that $\kappa_{\rm eff}\to 0$ and $\eta=\rho^*\pi \kappa_{\rm eff}/4$).
The agreement for the transition densities is
again very good between the theory and simulation. To see the effect of out-of-plane
orientational freedom it is worth plotting the IN densities of 2D hard ellipse system and those
of freely rotating hard oblate monolayer together (see Fig. \ref{fig7}). It can bee seen that the
simulation does not show a substantial difference between the two systems, while the theory
predicts that the out-of-plane freedom destabilizes the nematic order, i.e. the IN density curve
is shifted into the direction of higher densities. In the simulation the nematic phase evolves
from a very ordered planar phase (isotropic in the XY plane), where the effective aspect ratio
is almost identical with the aspect ratio of the ellipsoids (see Fig. \ref{fig5}), i.e. the 
corresponding
hard ellipse system is almost monodisperse. This is not the case in the theory, where the
transition happens at such densities, where the corresponding hard ellipse system is still
polydisperse with larger shape anisotropy ($\kappa_{\rm eff}$) than $\kappa$ . As a result the 
effect of out-of-plane
orientational freedom is more pronounced in the theory than in the simulation. This shows
that the theory exaggerates the effect of out-of-plane orientational freedom on the N-BN
phase transition.

\begin{figure}
\begin{center}
\includegraphics[width=0.8\linewidth,angle=0]{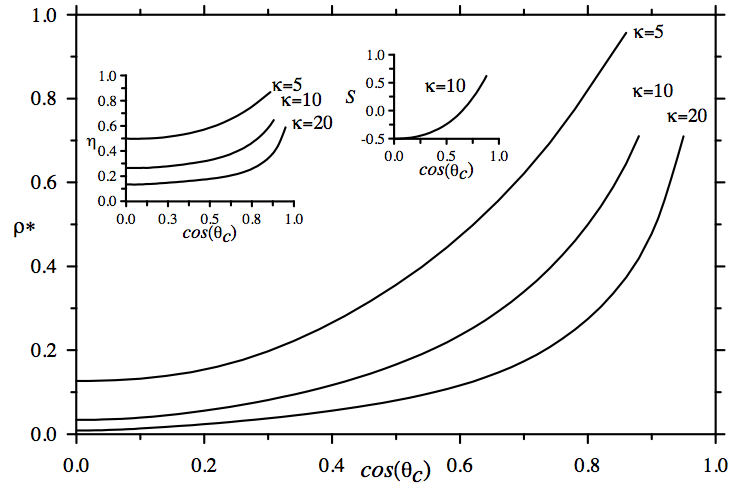}
\includegraphics[width=0.8\linewidth,angle=0]{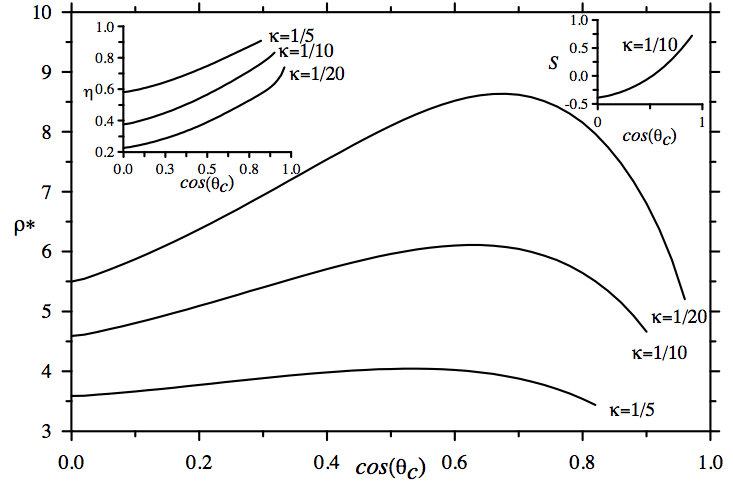}
\caption{
Monolayers of orientationally restricted hard ellipsoids: N-BN transition density vs.
$\cos\theta_c$. The upper panel is for prolate shapes while the lower one for oblate shapes. Insets
show the $\cos\theta_c$ dependence of the surface coverage and the out-of-plane order parameter at
the N-BN ordering transition. The curves are the results of PL theory.
}
\label{fig8}
\end{center}
\end{figure}

Finally we show how the IN transition properties change with switching on the out of
plane orientational freedom through the varying limiting polar angle ($\theta_c$), which is between 0
and $\pi/2$ for both prolate and oblate shapes. If $\theta_c=\pi/2$ ($\cos \theta_c=0$) the 
prolate ellipsoids on
the XY plane behave like the system of hard ellipses, i.e. they undergo a 2D IN phase
transition with increasing density (see Fig. \ref{fig8}). However this phase transition is 
destabilized with decreasing the limiting angle ($\theta_c$), which corresponds to increasing 
$\cos\theta_c$. As the
orientational window is widened (decreasing $\theta_c$), the prolate ellipsoids tend to minimize their
occupied area on the XY surface, which results in less anisotropic in-plane ellipses and higher
transition densities for all studied aspect ratios. In addition to this, the in-plane order
transforms continuously into out-of-plane order, i.e. S becomes positive with $\cos\theta_c$ 
(see Fig. \ref{fig8}). Hence, the 2D IN transitions turn into N-BN transitions for confined 
prolates which are
lost for the free rotating case. The surface coverage (packing fraction) curves show very
clearly that the occupied area on the XY surface increases enormously with decreasing the
polar angle restriction, i.e. the 2D IN phase transition is destabilized with decreasing $\theta_c$ . 
The value of $\theta_c$ where the orientational ordering transition is preempted by the positional one
(freezing transition) cannot be determined with the present PL theory. Interestingly, MC
simulation shows that the IN density is practically not affected by the value of limiting polar
angle, but the transition disappears at values of $\theta_c$ close to zero. Note that the IN transition
density curves move into the direction of lower densities with increasing shape anisotropy
because the occupied area of the ellipsoids becomes larger with increasing $\kappa$ as 
$\sigma_{\parallel}>\sigma_{\perp}$. The
case of oblate ellipsoids is different because $\theta_c=\pi/2$ corresponds to the freely rotating limit,
while $\theta_c=0$ is the hard disk limit. Starting from the freely rotating case one can see that the
IN transition density shows a maximum at an intermediate value of $\theta_c$ , which can be
attributed to the combined effect of decreasing in-plane shape anisotropy and the increasing
intersected area of the ellipsoids with decreasing $\theta_c$ . The N-BN transition packing fraction
behaves simply since it is always an increasing function of $\cos\theta_c$, i.e the nematic phase is
destabilized. This means that the nematic-biaxial nematic curve and the isotropic-solid curves
must cross each other at a threshold value of $\cos\theta_c$ making the nematic order metastable and
producing a plastic solid (an orientationally disordered solid). This plastic solid would
eventually turn into an ordered solid at higher densities. However, this value of $\cos\theta_c$ 
has not
been searched by the simulation and theory, because our study focuses only on the
orientational ordering. Finally, the $S$ order parameter shows that the system undergoes a
structural change from in-plane order into out-of-plane order at the N-BN transition. Our
results show that the effect of out-of-plane orientational freedom is to minimize the occupied
area on the surface which allows the most efficient packing of the ellipsoids on the XY plane.

\section{Conclusions}

We have examined the orientational ordering properties of the monolayer of hard
ellipsoids using replica exchange Monte Carlo simulation method and the Parsons-Lee density
functional theory. We have found that both the shape anisotropy and the orientational
restriction affect substantially the orientational order on the plane.
The subtle interplay
between orientational and packing entropies results in different structures and phase behaviors
for prolate and oblate shaped ellipsoids. The main driving force of the ordering behavior is to
minimize the intersected area between the particle and the surface to realize the close packing
structure with increasing surface density. The minimal intersected area is a disk for prolate,
while it is an ellipse for oblate ellipsoids. Highly packed structures can be achieved with only
out-of-plane (in-plane) ordering for prolate (oblate) ellipsoids, i.e. prolate ellipsoids prefer to
order along the normal of the plane, while oblate ellipsoids like to order in the plane.
The monolayer of prolate ellipsoids with very small out-of-plane orientational freedom
behaves almost identically to the system of 2D hard ellipses, i.e. it form 2D isotropic and
nematic phases. The gradual rise of the out-of-plane freedom, through decreasing $\theta_c$ , allows
the ellipsoid particle to lean out from the plane and to decrease the intersected area with the
plane. This involves less anisotropic shape in the interactions and destabilization of the biaxial
nematic phase (2D nematic) with respect to uniaxial nematic one (2D isotropic). At full
orientational freedom ($0<\theta<\pi$) prolate ellipsoids do not form an isotropic phase, but they
are ordered along the normal of the confining plane even at very low densities and behave
similarly to the 2D system of hard disks at high densities.

The monolayer of oblate ellipsoids without out-of-plane freedom is identical to the 2D
system of hard disks. The gradually increasing freedom in polar angle ($\theta$) allows the oblate
ellipsoid to decrease its intersected area with the confining plane through leaning out from the
plane. This makes the ordering planar for the main symmetry axis of the ellipsoid. However
the anisotropic interactions between the tilted ellipsoids give rise to an additional in-plane
order, i.e. the phase is biaxial nematic, which is due to the excluded area gain coming from
the in-plane ordering of elliptical intersections. In the freely rotating case the oblate ellipsoids
form a planar nematic phase at low densities and a biaxial nematic one at high densities.
Interestingly, the IN density of 2D ellipses is almost identical to that of freely rotating
ellipsoids confined to a plane.
In summary the out-of-plane orientational freedom stabilizes the in-plane nematic
ordering in the monolayer of oblate ellipsoids, while the opposite happens in the monolayer of
prolate ellipsoids. This is congruent with the 2D ellipse and 2D disk phase diagrams, which
are obtained as limiting cases in this study. In our simple model we have neglected the effect
of out-of-plane positional freedom which we expect to have a relatively small effect on the
system behavior. On the other hand, not including a soft wall-particle interaction may lead to
deviations from experimental set-ups where capillary phenomena are frequently important.
No doubt, the inclusion of this interaction would substantially increase the computation
burden of the problem. We leave this issue for future studies.

\acknowledgements

SV acknowledges the financial support of the Hungarian State and the European
Union under the TAMOP-4.2.2.A-11/1/KONV-2012-0071. Grant FIS2013-47350-C5-1-R
from Ministerio de Educaci\'on y Ciencia of Spain is also acknowledged.

\end{document}